\documentclass[a4paper,UKenglish,cleveref, autoref, thm-restate]{lipics-v2021}
\pdfoutput=1

\title{Type Theory with Explicit Universe Polymorphism
(revised and extended version)} 

\author{Marc Bezem}{University of Bergen, Norway}{Marc.Bezem@uib.no}%
{https://orcid.org/0000-0002-7320-1976}{}

\author{Thierry Coquand}{University of Gothenburg, Sweden} {Thierry.Coquand@cse.gu.se}%
{https://orcid.org/0000-0002-5429-5153}{}

\author{Peter Dybjer}{Chalmers University of Technology, Sweden}{peterd@chalmers.se}{https://orcid.org/0000-0003-4043-5204}{}

\author{Mart\'{\i}n Escard\'o}{University of Birmingham, UK}{m.escardo@bham.ac.uk}{https://orcid.org/0000-0002-4091-6334}{}

\authorrunning{M. Bezem and T. Coquand and P. Dybjer and M. Escard\'o}

\Copyright{Marc Bezem and Thierry Coquand and Peter Dybjer and Mart\'{\i}n Escard\'o} 



\category{} 

\relatedversion{} 

\hideLIPIcs
\nolinenumbers

\EventEditors{John Q. Open and Joan R. Access}
\EventNoEds{2}
\EventLongTitle{42nd Conference on Very Important Topics (CVIT 2016)}
\EventShortTitle{CVIT 2016}
\EventAcronym{CVIT}
\EventYear{2016}
\EventDate{December 24--27, 2016}
\EventLocation{Little Whinging, United Kingdom}
\EventLogo{}
\SeriesVolume{42}
\ArticleNo{23}

\newcommand{\eraser}[1]{}
\newcommand{\conv}{=}
\newcommand{\Id}{\mathsf{Id}}
\newcommand{\Eq}{\mathsf{Eq}}
\newcommand{\id}{\mathsf{id}}
\newcommand{\NN}{\mathsf{N}}

\newcommand{\Nat}{\mathbb{N}}
\newcommand{\UU}{\mathsf{U}}
\newcommand{\JJ}{\mathsf{J}}
\newcommand{\Type}{\mathsf{Type}}
\newcommand{\AgdaLevel}{\mathsf{Level}}
\newcommand{\Level}{\mathsf{level}}

\newcommand{\ZERO}{\mathsf{0}}
\newcommand{\SUCC}{\mathsf{S}}
\newcommand{\valid}{\mathsf{valid}}
\newcommand{\type}{\mathsf{type}}

\newcommand{\lam}[1]{{\langle}#1{\rangle}}
\newcommand{\mylam}[3]{\lambda_{#1:#2}#3}
\newcommand{\mypi}[3]{\Pi_{#1:#2}#3}
\newcommand{\Upi}[3]{\Pi^{#1}\,#2\,#3}
\newcommand{\mysig}[3]{\Sigma_{#1:#2}#3}
\newcommand{\Usig}[3]{\Sigma^{#1}\,#2\,#3}
\newcommand{\app}[2]{{#1\,#2}} 
\newcommand{\Sapp}[1]{\sapp{\SUCC}{#1}}
\newcommand{\sapp}[2]{{#1(#2)}} 
\newcommand{\Idapp}[3]{\sapp{\Id}{#1,#2,#3}}
\newcommand{\Idnapp}[4]{\sapp{\Id^#4}{#1,#2,#3}}
\newcommand{\NRapp}[4]{\sapp{\RR}{#1,#2,#3,#4}}
\newcommand{\Rfapp}[2]{\sapp{\refl}{#1,#2}}
\newcommand{\Japp}[6]{\sapp{\JJ}{#1,#2,#3,#4,#5,#6}}
\newcommand{\RR}{\mathsf{R}}

\newcommand{\Group}{\mathsf{Group}}

\newcommand{\T}{\mathsf{T}}

\newcommand{\idtoeq}{\mathsf{idtoeq}}

\newcommand{\Equiv}{\mathsf{Equiv}}
\newcommand{\isContr}{\mathsf{isContr}}
\newcommand{\ua}{\mathsf{ua}}
\newcommand{\UA}{\mathsf{UA}}

\newcommand{\set}[1]{\{#1\}}

\newcommand{\refl}{\mathsf{refl}}

\newcommand{\AFu}{\mathcal{A}}
\newcommand{\Fu}{\mathit{Fu}}

\begin{document}

\maketitle

\begin{abstract}
The aim of this paper is to refine and extend proposals by Sozeau and Tabareau and by Voevodsky for universe polymorphism in type theory. In those systems judgments can depend on explicit constraints between universe levels. We here present a system where we also have products indexed by universe levels and by constraints. Our theory has judgments for internal universe levels, built up from level variables by a successor operation and a binary supremum operation, and also judgments for equality of universe levels.
\end{abstract}

\section{Introduction}\label{sec:intro}

The system of simple type theory,
as introduced by Church \cite{church:formulation},
is elegant and forms the basis of several proof assistants.
However, it has some unnatural limitations:
it is not possible in this system to talk
about an arbitrary type or about an arbitrary structure.
For example, it is not possible to form the collection of all groups
as needed in category theory. In order to address these limitations,
Martin-L\"of \cite{ML71,ML71a} introduced a system with a type $V$ of all types.
A function $A\rightarrow V$ in this system can then be seen as a family of types
over a given type $A$. It is natural in such a system to refine
the operations exponential and cartesian product in simple type theory
to operations of dependent products and sums.
After the discovery of Girard's paradox \cite{Girard71},
Martin-L\"of \cite{ML72} introduced a distinction between
{\em small} and {\em large} types, similar to the distinction introduced
in category theory between large and small sets,
and the type $V$ became the (large) type of small types.
The name ``universe'' for such a type was chosen in analogy with the notion of
universe introduced by Grothendieck to represent category theory in set theory.

Later, Martin-L\"of \cite{martinlof:predicative} introduced a countable sequence of universes
$$
\UU_0 : \UU_1 : \UU_2 : \cdots
$$
We refer to the indices $0, 1, 2, \ldots$ as {\em universe levels}.

Before the advent of univalent foundations, most type theorists expected
only the first few universe levels to be relevant in practical formalisations.
One thought that it might be feasible for a user of
type theory to explicitly assign universe levels to their types and
then simply add updated versions of earlier
definitions when they were needed at different levels.
However, the number of copies of definitions does not only grow with the level,
but also with the number of type arguments in the definition of a type former.
(The latter growth can be exponential!)

To deal with this, Huet \cite{Huet87} introduced a specific form of
universe polymorphism that allowed the use of $\UU:\UU$
on the condition that each occurrence of $\UU$ can be disambiguated
as $\UU_i$ in a consistent way.
This approach has been followed by Harper and Pollack \cite{HarperP91} and
in Coq \cite{coq:general}.
These  approaches to \emph{implicit} universe polymorphism are, however,
problematic with respect to modularity. As pointed out in \cite{Courant02,Simpson04}:
one can prove $A\rightarrow B$ in one file, and $B\rightarrow C$ in
another file, while $A\rightarrow C$ is not valid.

Leaving universe levels implicit also causes practical problems,
since universe level disambiguation can be a costly operation,
slowing down type-checking significantly.
Moreover, so-called universe inconsistencies can be hard to explain to the user.

In order to cope with these issues, Courant \cite{Courant02}
introduced explicit universe levels,
with a supremum operation (see also Herbelin \cite{herbelin05}).
Explicit universe levels are also present in Agda \cite{agda-manual} and
Lean \cite{moura:lean,Carneiro19}.
However, whereas Courant has
universe level \emph{judgments}, Agda has a \emph{type} of
universe levels, and hence supports the formation of level-indexed products.

With the advent of Voevodsky's univalent foundations,
the need for universe polymorphism has only increased.
One often wants to prove theorems uniformly for arbitrary universes. These theorems may depend on several universes and there may be constraints on the level of these universes.
In response to this Voevodsky  \cite{VV} and Sozeau and Tabareau \cite{SozeauTabareau:coq} proposed type theories parameterized by
(arbitrary but fixed) universe levels and constraints.

The \emph{univalence axiom} states that for any two types $X,Y$ the canonical map
$$
\idtoeq_{X,Y} : (X=Y)\to (X\simeq Y)
$$
is an equivalence.
Formally, the univalence axiom is an axiom scheme which is added to
Martin-Löf type theory.
If we work in Martin-Löf type theory with a countable tower of universes,
each type is a member of some universe $\UU_n$.
Such a universe $\UU_n$ is {\em univalent} provided for all $X,Y : \UU_n$ the
canonical map $\idtoeq_{X,Y}$ is an equivalence.
Let $\UA_n$ be the type expressing the univalence of $\UU_n$, and let
$\ua_n : \UA_n$ for $n = 0,1,\ldots$ be a sequence of constants postulating
the respective instances of the univalence axiom.
We note that $X = Y : \UU_{n+1}$ and $X\simeq Y : \UU_n$ and
hence $\UA_n : \UU_{n+1}$. We can express the universe polymorphism of these judgments internally in all of the above-mentioned systems by quantifying over universe levels, irrespective of having universe level judgments or a type of universe levels.

To be explicit about universes can be important, as shown by Waterhouse~\cite{waterhouse:sheaves,chambert-loir:universes-matter}, who gives an example of a large presheaf with no associated sheaf. A second example is the fact that the embedding
 $\Group(\UU_n)\rightarrow \Group(\UU_{n+1})$ of the type of groups in a universe $\UU_n$ into that of the next universe $\UU_{n+1}$ is not an equivalence. That is, there are more groups in the next universe~\cite{bcde:largegroup}.

 We remark that universes are even more important in a predicative framework
than in an impredicative one, for uniform proofs and modularity.
Consider for example the formalisation of real numbers as Dedekind cuts,
or domain elements as filters of formal neighbourhoods. Both belong to $\UU_1$ since they are properties of elements in~$\UU_0$.
However, even in a system using an impredicative universe of propositions,
such as the ones in \cite{Huet87,moura:lean}, there is a need for
definitions parametric in universe levels.

\paragraph*{Terminology}
Following Cardelli \cite{Cardelli87}, we distinguish between
{\em implicit} and {\em explicit} polymorphism:
\begin{quotation}
  Parametric polymorphism is explicit when parametrization is obtained by
explicit type parameters in procedure headings, and corresponding explicit
applications of type arguments when procedures are called $\dots$
Parametric polymorphism is called implicit when the above type parameters and type applications are not admitted, but
types can contain type variables which are unknown, yet to be determined, types.
\end{quotation}

\paragraph*{Motivation}
Many substantial Agda developments make essential use of \emph{explicit} universe polymorphism with successor and finite suprema. Examples include the Agda standard library~\cite{agda:stdlib}, the cubical Agda library~\cite{cubical:agda}, 1Lab~\cite{agda:1lab}, the Agda-HoTT library~\cite{hott:agda}, agda-unimath~\cite{agda:unimath}, TypeTopology~\cite{TypeTopology},
HoTT-UF-in-Agda~\cite{hott:uf:in:agda} (Midlands Graduate School 2019 lecture notes).

The original motivation for this work was to formalise the type theory of Agda, including explicit universe polymorphism. In doing that, we found ourselves modifying Agda's treatment of universes as follows:
\begin{itemize}
\item We have universe level \emph{judgments}, like Courant~\cite{Courant02}, instead of a \emph{type} of universe levels, like Agda.

 \item We add the possibility of expressing \emph{explicit} universe
 level constraints. This is not only more general but also arguably
 gives a more natural way of expressing types involving universes.

 \item We do not require a first universe level zero, so that every definition that involves universes is polymorphic.

 \item We include a $\Type$ judgment, which does not refer to universes, as in Martin-L\"of~\cite{martinlof:hannover}.

\end{itemize}

Our resulting type theory is orthogonal to the presence or absence of cumulativity. In the body of the paper, we treat universes \`a la Tarski, but we also give an appendix with a version \`a la Russell.

We have checked that the lecture notes~\cite{hott:uf:in:agda} on HoTT/UF, which include 9620 lines of Agda without comments, can be rewritten without universe level zero. We believe, based on what we learned from this experiment, that the above Agda developments could also be rewritten in this way. Experience with these Agda developments suggest that a \emph{type} for levels in Agda could be replaced by level judgments in practice. The fact that levels form a type in Agda automatically allows for nested universal quantification over levels, which we instead add explicitly to our type theory.

\paragraph*{Summary of main contributions}
Like Courant, we present a type theory with
universe levels  and universe level equations as \emph{judgments}. Moreover, we don't restrict the levels to be natural numbers. Instead we just assume that they form a sup-semilattice with an inflationary endomorphism. In this way all levels are built up from level variables by a successor operation and a binary supremum operation. Unlike most other systems, we do not have a level constant $0$ for the first universe level. Thus all types involving universes depend on level variables; they are {\em universe polymorphic}.

Furthermore, we make the polymorphism fully explicit in the sense of Cardelli
by adding \emph{level-indexed products}. In this way we regain some of
the expressivity Agda gets from having a \emph{type}
$\AgdaLevel$ of universe levels.
Finally, we present a type theory with constraints as \emph{judgments} similar to the ones by Sozeau and Tabareau \cite{SozeauTabareau:coq} and Voevodsky \cite{VV} but extended with \emph{constraint-indexed products}.

\paragraph*{Plan}
In Section~\ref{sec:basic} we display rules for a basic version of dependent
type theory with $\Pi, \Sigma, \NN$, and an identity type former $\Id$.

In Section~\ref{sec:external} we explain how to add an externally indexed
sequence of universes
$\UU_n, \T_n~(n\in\Nat)$ \`a la Tarski, without cumulativity rules.
In Appendix~\ref{appendix:1} we present a system with cumulativity,
and in Appendix~\ref{appendix:2} we present a system \`a la Russell.

In Section~\ref{sec:internal} we introduce a notion of universe level,
and let judgments depend not only on a context of ordinary variables,
but also on level variables $\alpha, \ldots, \beta$.
This gives rise to a type theory with level polymorphism, which we call
``ML-style'' as long as we do not bind level variables.
We then extend this theory with level-indexed products of
types $[\alpha]A$ and corresponding abstractions $\lam{\alpha}A$
to give full level polymorphism.

In Section~\ref{sec:constraints} we extend the type theory in
Section~\ref{sec:internal} with constraints (lists of equations between level expressions). Constraints can now appear as assumptions in hypothetical judgments. Moreover, we add
constraint-indexed products of types $[\psi]A$ and corresponding
abstractions $\lam{\psi}A$. This goes beyond the systems of Sozeau and Tabareau \cite{SozeauTabareau:coq} and Voevodsky \cite{VV}.
In Section~\ref{sec:related} we
compare our type theory with Voevodsky's and Sozeau-Tabareau's
and briefly discuss some other approaches.
Finally, in Section~\ref{sec:future} we outline future work.

Appendix~\ref{appendix:3} has been added in the Fall of 2024
and contains corrections and extensions.

\section{Rules for a basic type theory}\label{sec:basic}

We begin by listing the rules for a basic type theory
with $\Pi, \Sigma, \NN,$ and $\Id$. A point of departure is
the system described by Abel et al.\ in \cite{abel18}, since a
significant part of the metatheory of this system has been formalized in Agda.
This system has $\Pi$-types, $\NN$ and one universe.
However, for better readability we use named variables instead
of de Bruijn indices. We also add $\Sigma$ and $\Id$, and,
in the next sections, a tower of universes.

The judgment $\Gamma\vdash$ expresses that $\Gamma$ is a context.
The judgment $\Gamma\vdash A$ expresses that $A$ is a type in context $\Gamma$.
The judgment $\Gamma\vdash a:A$ expresses that $A$ is a type
and $a$ is a term of type $A$ in context $\Gamma$. The rules are given in \cref{fig:context}.

\begin{figure}[h!]
$$
\frac{}{()\vdash}~~~~~~~
\frac{\Gamma\vdash A}{\Gamma,x:A\vdash}~(x~\text{fresh})~~~~~~
\frac{\Gamma\vdash}{\Gamma\vdash x:A}~(x\!:\! A~\text{in}~\Gamma)
\belowdisplayshortskip 0pt
$$
\caption{Rules for context formation and assumption}\label{fig:context}
\end{figure}

We may also write $A~\type~(\Gamma)$ for $\Gamma\vdash A$,
and may omit the global context $\Gamma$,
or the part of the context that is the same for all hypotheses and for the
conclusion of the rule.
Hypotheses that could be obtained from other
hypotheses through inversion lemmas are often left out,
for example, the hypothesis $A~\type$ in the first rule for $\Pi$ and $\Sigma$
in \cref{fig:PiSig}.

\begin{figure}[h!]
$$
\frac{B~\type~(x:A)}{\mypi{x}{A}{B}~\type}~~~~~~~~~
\frac{b:B~(x:A)}{\mylam{x}{A}{b}:\mypi{x}{A}{B}}~~~~~~~~
\frac{c:\mypi{x}{A}{B}~~~~~~a:A}
     {\app{c}{a}:B(a/x)}
$$
$$
\frac{B~\type~(x:A)}{\mysig{x}{A}{B}~\type}~~~~~~~~~
\frac{a:A~~~~~~b:B(a/x)}{(a,b):\mysig{x}{A}{B}}~~~~~~~~
\frac{c:\mysig{x}{A}{B}}{c.1:A}~~~~~~~
\frac{c:\mysig{x}{A}{B}}{c.2:B(c.1/x)}
\belowdisplayshortskip -1pt
$$
  \caption{Rules for $\Pi$ and $\Sigma$}\label{fig:PiSig}
\end{figure}

We write $\conv$ for definitional equality (or conversion).
The following rules express that conversion is an equivalence
relation and that judgments are invariant under conversion.
The rules are given in Figures~\ref{fig:conversion} and~\ref{fig:convPiSig}.

\begin{figure}[h!]
$$
\frac{ a:A~~~~~~ A~ \conv~ B}{ a:B}~~~~~~~~~
\frac{ a ~\conv~a':A~~~~~~ A  ~\conv~ B}{ a ~\conv~a':B}
$$
$$
\frac{A~=~B~~~~~A~=~C}{B~=~C}~~~~~~~~~\frac{A~\type}{A~=~A}~~~~~~~~~
\frac{a~=~b:A~~~~~a~=~c:A}{b~=~c:A}~~~~~~~~~\frac{a:A}{a~=~a:A}
\belowdisplayshortskip 1pt
$$
  \caption{General rules for conversion}\label{fig:conversion}
\end{figure}

\begin{figure}[h!]
$$
\frac{A~=~A'~~~~~~B~=~B'~(x:A)}{\mypi{x}{A}{B}~=~\mypi{x}{A'}{B'}}~~~~~~~~
\frac{c~=~c':\mypi{x}{A}{B}~~~~~~a~=~a':A}{c~a~=~c'~a':B(a/x)}
$$
$$
\frac{b:B~(x:A)~~~~~~~~ a:A}{ \mylam{x}{A}{b}~a  ~\conv~ b(a/x):B(a/x)}
~~~~~~~
\frac{f~x = g~x:B~(x:A)}{ f = g : \mypi{x}{A}{B}}
$$
$$
\frac{A~=~A'~~~~~~B~=~B'~(x:A)}{\mysig{x}{A}{B}~=~\mysig{x}{A'}{B'}}~~~~~~~~
\frac{c~=~c':\mysig{x}{A}{B}}{c.1~=~c'.1:A}~~~~~~~~
\frac{c~=~c':\mysig{x}{A}{B}}{c.2~=~c'.2:B(c.1/x)}~~~~~~~~
$$
$$
\frac{a:A~~~~~b:B(a/x)}{ (a,b).1 ~\conv~ a:A}~~~~~~~~~~
\frac{a:A~~~~~b:B(a/x)}{ (a,b).2 ~\conv~ b:B(a/x)}~~~~~~~~~~
\frac{c.1~=~ c'.1:A~~~~~~c.2~=c'.2:B(c.1/x)}{ c~=~ c' : \mysig{x}{A}{B}}
\belowdisplayshortskip 0pt
$$
  \caption{Conversion rules for $\Pi$ and $\Sigma$}\label{fig:convPiSig}
\end{figure}

By now we have introduced several parametrized syntactic constructs
for types and terms, such as $\mypi{x}{A}{B}$,
$\mylam{x}{A}{b}$, $\app{c}{a}$, $(a,b).2$.
Conversion rules for $\Pi$ and $\Sigma$ were given in \cref{fig:convPiSig}.
and those rules imply that $=$ is a congruence.%
(Some cases of congruence are subtle. Exercise:
show congruence of $=$ for $\mylam{x}{A}{b}$ and $(a,b)$.)
In the sequel we will tacitly assume the inference rules
ensuring that $=$ is a congruence for all syntactic constructs
that are to follow.

We now introduce the type of natural numbers $\NN$ with
the usual constructors $\ZERO,\SUCC$ and eliminator $\RR$,
as an example of an inductive data type.
Rules with the same hypotheses are written as one rule with
several conclusions.
The rules are given in \cref{fig:typeN}.

\begin{figure}[b]
$$
\frac{}{\NN~\type~~~~\ZERO:\NN}~~~~~
\frac{n:\NN}{\Sapp{n} : \NN}~~~~~~
\frac{P~\type~(x:\NN)~~~~a:P(\ZERO/x)~~~~~
g:\mypi{x}{\NN}{(P\to P(\Sapp{x}/x))}}
{\NRapp{P}{a}{g}{\ZERO} = a: P(\ZERO/x) }
$$
$$
\frac{P~\type~(x:\NN)~~~~a:P(\ZERO/x)~~~~~
g:\mypi{x}{\NN}{(P\to P(\Sapp{x}/x))}~~~~~n:\NN}
{\NRapp{P}{a}{g}{n}:P(n/x)~~~~~~\NRapp{P}{a}{g}{\Sapp{n}} = g~n~\NRapp{P}{a}{g}{n}: P(\Sapp{n}/x) }
$$
  \caption{Rules and conversion rules for the datatype $\NN$}\label{fig:typeN}
\end{figure}

We also add identity types $\Idapp{A}{a}{a'}$ for all $A~\type$,
$a:A$ and $a':A$, with constructor $\Rfapp{A}{a}$ and (based) eliminator
$\Japp{A}{a}{C}{d}{a'}{q}$. The rules are given in \cref{fig:typeId}.

\begin{figure}[t]
$$
\frac{A~\type ~~~~ a:A ~~~~ a':A}{\Idapp{A}{a}{a'}~\type}~~~~~~~
\frac{a:A}{\Rfapp{A}{a}:\Idapp{A}{a}{a}}
$$
$$
\frac{a:A~~~~C~\type~(x:A,p:\Idapp{A}{a}{x})~~~~d:C(a/x,\Rfapp{A}{a}/p)
~~~~a':A~~~~q:\Idapp{A}{a}{a'}}
{\Japp{A}{a}{C}{d}{a'}{q}: C(a'/x,q/p) ~~~~~~~~
 \Japp{A}{a}{C}{d}{a}{\Rfapp{A}{a}} = d : C(a/x,\Rfapp{A}{a}/p)}
\belowdisplayshortskip 0pt
$$
  \caption{Rules and conversion rule for identity types}\label{fig:typeId}
\end{figure}

In this basic type theory we can define, for example,
$\isContr(A) := \mysig{a}{A}{\mypi{x}{A}{\Id(A,a,x)}}$
for $A~\type$, expressing that $A$ is contractible.
If also $B~\type$, we can define $\Equiv(A,B) :=
\mysig{f}{A\to B}{\mypi{b}{B}{\isContr(\mysig{x}{A}{\Id(B,b,f(x))})}}$,
which is the type of equivalences from $A$ to $B$. This example
will also be used later on.

\section{Rules for an external sequence of universes}\label{sec:external}

We present an external sequence of universes of codes of types, together
with the decoding functions. (We do not include rules for cumulativity here, leaving them for Appendix~\ref{appendix:1}.)
The rules are given in \cref{fig:typeU}.

\begin{figure}[h!]
$$
\frac{}{\UU_m~\type}~~~~~~
\frac{A:\UU_{m}}{\T_{m}(A)~\type}~~~~~~
\frac{}{\UU^{n}_{m}:\UU_{n}~~~~~~\T_{n}({\UU^{n}_{m}}) = \UU_{m}}{~(n>m)}
\belowdisplayshortskip 0pt
$$
  \caption{Rules and conversion rules for all universes $\UU_m$ and their codes $\UU^{n}_{m}~(n>m)$}\label{fig:typeU}
\end{figure}

Here and below $m$ and $n$, as super- and subscripts of $\UU$ and $\T$,
are \emph{external} natural numbers, and $n \vee m$ is the
maximum of $n$ and $m$. This means, for example, that $\UU_m~\type$ is
a \emph{schema}, yielding one rule for each $m$.

Next we define how $\Pi, \Sigma, \NN,$ and $\Id$ are ``relativized'' to
codes of types, and how they are decoded, in Figures~\ref{fig:PiSigU} and~\ref{fig:NIdU}.

\begin{figure}[h!]
$$
\frac{A:\UU_{n}~~~~~~B:\T_{n}(A)\rightarrow \UU_{m}}
     {\Pi^{n,m} A B:\UU_{n\vee m}~~~~~~~~~
      \T_{n\vee m}~(\Pi^{n,m} A B) = \mypi{x}{\T_{n}(A)}{ \T_{m}(B~x)}}
$$
$$
\frac{A:\UU_{n}~~~~~~B:\T_{n}(A)\rightarrow \UU_{m}}
     {\Sigma^{n,m} A B:\UU_{n\vee m}~~~~~~~~~
     \T_{n\vee m}~(\Sigma^{n,m} A B) = \mysig {x}{\T_{n}(A)}{ \T_{m}(B~x)}}
\belowdisplayshortskip -1pt
$$
  \caption{Rules and conversion rules for $\Pi$ and $\Sigma$ for codes of types}\label{fig:PiSigU}
\end{figure}

\begin{figure}[h!]
$$
\frac{}{\NN^{n}:\UU_{n}}~~~~~~\frac{}{\T_{n}(\NN^{n}) = \NN}
$$
$$
\frac{A:\UU_n~~~~a_0:\T_n(A)~~~~a_1:\T_n(A)}
{\Idnapp{A}{a_0}{a_1}{n}:\UU_n ~~~~~~~~~ \T_n(\Idnapp{A}{a_0}{a_1}{n}) = \Idapp{{\T_n(A)}}{a_0}{a_1} }
\belowdisplayshortskip 1pt
$$
\caption{Rules and conversion rules for codes of $\NN$ and $\Id$}\label{fig:NIdU}
\end{figure}

In the following section we present a type theory with \emph{internal}
universe level expressions. This theory has finitely many inference rules.

 \section{A type theory with universe levels and polymorphism }\label{sec:internal}

The problem with the type system with an external sequence of universes
is that we have to \emph{duplicate} definitions that follow
the same pattern. For instance, we have the identity function
$$
\id_n := \mylam{X}{\UU_n}{\mylam{x}{\T_n(X)}{x}} : \mypi{X}{\UU_n}{\T_n(X)\rightarrow \T_n(X)}
$$
This is a schema that
may have to be defined (and type-checked) for several $n$.
We address this issue by introducing \emph{universe level}
expressions: we write $\alpha,\beta,\dots$
for \emph{level variables}, and $l,m,\dots$ for
\emph{level expressions} which are built from level variables
by suprema $l \vee m$ and the next level operation $l^+$.
Level expressions form a sup-semilattice $l\vee m$
with a next level operation $l^+$ such that $l \vee l^+ = l^+$
and $(l\vee m)^+ = l^+\vee m^+$. (We don't need a $0$ element.)
We write $l\leqslant m$ for $l\vee m = m$ and $l<m$ for $l^+\leqslant m$.
See \cite{bezem-coquand:lattices} for more details.

We have a new context extension operation that adds a fresh level variable
$\alpha$ to a context, a rule for assumption, and typing rules
for level expressions, in \cref{fig:contextL}.

\begin{figure}[h!]
$$
\frac{\Gamma\vdash}{\Gamma,\alpha~\Level\vdash}~(\alpha~\text{fresh})~~~~~~
\frac{\Gamma\vdash}{\Gamma\vdash\alpha~\Level}(\alpha~\text{in}~\Gamma)~~~~~~
\frac{l~\Level~~~~~m~\Level}{l\vee m~\Level}~~~~~~
\frac{l~\Level}{l^+~\Level}
\belowdisplayshortskip 0pt
$$
  \caption{Rules for typing level expressions, extending
  \cref{fig:context}}\label{fig:contextL}
\end{figure}

We also have level equality judgments $\Gamma\vdash l = m$
and want to enforce that judgments are invariant under level equality.
To this end we add the rule that $\Gamma\vdash l = m$
when $\Gamma\vdash l~\Level$ and $\Gamma\vdash m~\Level$ and
$l=m$ in the free sup-semilattice above with $\_^+$ and generators
(level variables) in $\Gamma$.

In the next section we will also consider \emph{hypothetical} level
equality judgments, i.e., we may have constraints in $\Gamma$,
quotienting the free sup-semilattice above.

We tacitly assume additional 
rules ensuring that level equality
implies definitional equality of types and terms.
It then follows from the rules of our basic type theory that
judgments are invariant under level equality: if $l=m$ and
${a(l/\alpha) : A(l/\alpha)}$, then ${a(m/\alpha) : A(m/\alpha)}$.

We will now add rules for internally indexed universes in \cref{fig:typeUl}.
Note that $l<m$ is shorthand for the level equality judgment
$m= l^+ \vee m$.
\begin{figure}[h!]
$$
\frac{l~\Level}{\UU_l~\type}~~~~~~
\frac{A:\UU_{l}}{\T_{l}(A)~\type}~~~~~~
\frac{l<m}{\UU^{m}_{l}:\UU_{m}~~~~~~\T_{m}({\UU^{m}_{l}}) = \UU_{l}}
\belowdisplayshortskip 0pt
$$
  \caption{Rules and conversion rule for universes $\UU_l$ and their codes}\label{fig:typeUl}
\end{figure}

The remaining rules are completely analogous to the rules in
\cref{fig:PiSigU} and \cref{fig:NIdU}
for externally indexed universes with external numbers replaced
by internal levels. (To rules without assumptions, such as the first two
in Fig.~\ref{fig:NIdU}, we need to add assumptions like $n~\Level$,
for other rules these assumptions can be obtained from inversion lemmas.)

 We expect that normalisation holds for this system.
 This would imply decidable type-checking.
 This would also imply that if $a : \NN$ in a context with only
 level variables, then $a$ is convertible to a numeral.

\paragraph*{Interpreting the level-indexed system in the system with externally indexed universes}

A judgment in the level-indexed system can be interpreted in the externally indexed system relative to an assignment $\rho$ of external natural numbers to level variables. We simply replace each level expression in the judgment by the corresponding natural number obtained by letting $l^+\,\rho = l\,\rho+1$ and $(l \vee m)\,\rho = \max(l\,\rho,m\,\rho)$.

\paragraph*{Rules for level-indexed products}

In Agda $\AgdaLevel$ is a type,
and it is thus possible to form level-indexed products of types as $\Pi$-types.
In our system this is not possible, since $\Level$ is not a type. Nevertheless, it is useful for modularity to be able to form level-indexed products. Thus we extend the system with the rules in \cref{fig:levindprod}.
\begin{figure}[h!]
$$
\frac{A~\type~(\alpha~\Level)}{[\alpha]A~\type}~~~~~~~
\frac{t:[\alpha]A~~~~~l~\Level}
     {t~l:A(l/\alpha)}~~~~~~~~~
\frac{u:A~(\alpha~\Level)}{\lam{\alpha}{u}: [\alpha]A}~~~~~
\frac{t~\alpha = u~\alpha:A~(\alpha~\Level)}{t = u:[\alpha]A}
$$
$$
\frac{u:A~(\alpha~\Level)~~~~~l~\Level}
{(\lam{\alpha}{u})~l = u(l/\alpha): A(l/\alpha)}
$$
  \caption{Rules and conversion rule for level-indexed products}%
  \label{fig:levindprod}
\end{figure}

In this type theory we can reflect, for example, $\isContr(A) :=
\mysig{a}{A}{\mypi{x}{A}{\Id(A,a,x)}}$ for $A~\type$ as follows.
In the context $\alpha\,\Level, A: \UU_\alpha$, define
\[
\isContr^\alpha(A) :=
\Usig{\alpha,\alpha}{A}{(\mylam{a}{\T_\alpha(A)}
{(\Upi{\alpha,\alpha}{A}{(\mylam{x}{\T_\alpha(A)}{\Id^\alpha(A,a,x)})})})}.
\]
Then $\T_\alpha (\isContr^\alpha(A)) = \isContr(\T_\alpha(A))$.
We can further abstract to obtain the following typing:
\[
\lam{\alpha}{\mylam{A}{\UU_\alpha}{\isContr^\alpha(A)}} :
[\alpha](\UU_\alpha \to \UU_\alpha).
\]
In a similar way we can reflect $\Equiv(A,B)$ for $A,B\,\type$ by defining
in context $\alpha\,\Level,\beta\,\Level$, $A: \UU_\alpha, B: \UU_\beta$
a term $\Eq^{\alpha,\beta}(A,B) : \UU_{\alpha\vee\beta}$
such that $\T_{\alpha\vee\beta}(\Eq^{\alpha,\beta}(A,B))=
\Equiv(\T_{\alpha}(A),\T_{\beta}(B))$.

An example that uses level-indexed products beyond the ML-style
polymorphism (provided by Sozeau and Tabareau and by Voevodsky)
is the following type which
expresses the theorem that univalence for universes of arbitrary level implies
function extensionality for functions between universes of arbitrary levels.
$$
([\alpha]\mathsf{IsUnivalent}\, \UU_\alpha)
\to [\beta][\gamma] \mathsf{FunExt}\, \UU_\beta\, \UU_\gamma
$$
In other words, {\em global} univalence implies {\em global} function
extensionality.

Since an assumption of global function extensionality can replace many assumptions of local function extensionality (provided by ML-style polymorphism), this can also give rise to shorter code, see the example {\tt Eq-Eq-cong'} in~\cite{hott:uf:in:agda}.

\section{A type theory with level constraints}\label{sec:constraints}

To motivate why it may be useful to introduce the notion of judgment relative to a list of constraints on universe levels, consider the following type in a system without cumulativity. (We use Russell style notation for readability, see Appendix~\ref{appendix:2} for the rules for the Russell style version of our system.)

$$
    \Pi_{A:\UU_l~{B}:{\UU_m}~{C}:{\UU_n}}
    {~~\Id\,\UU_{l \vee m}\, (A\times B)\,(C \times A)
    \to \Id\,\UU_{m \vee l} \, (B\times A)\,(C \times A)}
$$
This is well-formed provided $l \vee m = n \vee l$.
There are several independent solutions:
\begin{eqnarray*}
&&l = \alpha, m = \beta, n = \alpha \vee \beta\\
&&l = \alpha, m = \gamma \vee \alpha, n = \gamma\\
&&l = \beta \vee \gamma, m = \beta, n = \gamma\\
&&l = \alpha, m = \beta, n = \beta
\end{eqnarray*}
where $\alpha, \beta,$ and $\gamma$ are level
variables. It should be clear that there cannot be any {\em most general} solution, since this solution would have to
assign variables to $l,m,n$.

In a system with level constraints,
we could instead derive the (inhabited under $\UA$) type
$$
    \Pi_{A:\UU_\alpha~{B}:{\UU_\beta}~{C}:{\UU_\gamma}}
    {~~\Id\,\UU_{\alpha \vee \beta}\, (A\times B)\,(C \times A)
    \to \Id\,\UU_{\alpha \vee \gamma}\, (B\times A)\,(C \times A)}
$$
which is valid under the constraint
$\alpha \vee \beta = \alpha \vee \gamma$,
which captures all solutions simultaneously.

Without being able to declare explicitely such constraints,
one would instead need to write four separate definitions.

 Surprisingly, if we add a least level $0$ to the term levels (like in Agda) then there \emph{is} a most general solution,
 namely $l = \alpha\vee\beta\vee\delta,~m = \beta\vee\gamma,~n = \alpha\vee\gamma$,\footnote{We learnt this from
 Thiago Felicissimo, with a reference to the work \cite{FBB:impred2pred}.}
 since it can be seen as an instance of
 a Associative Commutative Unit Idempotent unification problem \cite{BaaderS94}.

 It is however possible to find equation systems which do not have a most general unifier, even with a least level $0$, using
 the next level operation. For instance, the system $l^+ = m\vee n$ does not have a most general unifier, using a reasoning
 similar to the one in \cite{FBB:impred2pred}.

\paragraph*{Rules for level constraints}

A constraint is an equation $l = m$, where $l$ and $m$ are level expressions.
Voevodsky \cite{VV} suggested to introduce universe levels with
constraints. This corresponds to mathematical practice:
for instance, at the beginning of the book \cite{giraud:cohom-non-abel},
the author introduces two universes $U$ and $V$ with the constraint
that $U$ is a member of $V$.
In our setting this will correspond to introducing two levels
$\alpha$ and $\beta$ with the constraint $\alpha<\beta$.

Note that $\alpha < \beta$ holds iff $\beta = \beta \vee\alpha^+$.
We can thus avoid declaring this constraint if we instead
systematically replace $\beta$ by $\beta\vee\alpha^+$.
This is what is currently done in the system Agda.
However, this is a rather indirect way to express what is
going on. Furthermore, the example at the beginning of this section
shows that this can lead to an artificial duplication of definitions.

Recall that we have in Section~\ref{sec:internal}, e.g., the rule
that $\UU^m_l:\UU_m$ if $l<m~\valid$, that is, if
$l<m$ holds in the free semilattice. 
In the extended system in this section, this typing rule also applies
when $l<m$ is implied by the constraints in the context $\Gamma$.
For instance, we have $\alpha^+\leqslant\beta$ in a context
with constraints $\alpha\leqslant\gamma$ and $\gamma^+\leqslant\beta$.

To this end we introduce a new context extension operation $\Gamma,\psi$
extending a context $\Gamma$ by a finite set of constraints $\psi$.
The first condition for forming $\Gamma,\psi$ is that all level variables
occurring in $\psi$ are declared in $\Gamma$. The second condition
is that the finite set of constaints in the extended context
$\Gamma,\psi$ is loop-free.
A finite set of constraints is {\em loop-free} if it does not
create a {\em loop}, i.e., a level expression $l$ such that $l<l$
modulo this set of constraints, see \cite{bezem-coquand:lattices}.

We also have a new judgment form $\Gamma\vdash\psi~\valid$ that expresses
that the constraints in $\psi$ hold in $\Gamma$, that is,
are implied by the constraints in $\Gamma$. If there are
no constraints in $\Gamma$, the judgment $\Gamma\vdash\set{l=m}~\valid$
amounts to the same as $\Gamma\vdash l=m$ in Section~\ref{sec:internal}.
Otherwise it means that the constraints in $\psi$ hold in the
sup-semilattice with $\_^+$ presented by $\Gamma$.

As shown in \cite{bezem-coquand:lattices}, $\Gamma\vdash\psi~\valid$
as well as loop-checking, is decidable in polynomial time.

Voevodsky \cite {VV} did not describe a mechanism to {\em eliminate}
universe levels and constraints. In Figure~\ref{fig:levindprod} we
gave rules for eliminating universe levels
and in Figure~\ref{fig:restriction} below we give rules
for eliminating universe level constraints.

\paragraph*{Rules for constraint-indexed products}
We introduce a ``restriction'' or ``constraining'' operation with
the rules in \cref{fig:restriction}.
\begin{figure}[h!]
$$
\frac{A~\type~~(\psi)}{[\psi]A~\type}~~~~~~~~~
\frac{{t}:A~~(\psi)}{\lam{\psi}{t}:[\psi]A}~~~~~~~~~
\frac{\psi~\valid}{[\psi]A = A}~~~~~~~~~
\frac{\psi~\valid}{\lam{\psi}{t} = t}
\belowdisplayshortskip 1pt
$$
 \caption{Rules for constraining}%
  \label{fig:restriction}
\end{figure}

Here is a simple example of the use of this system.
In order to represent set theory in type
theory, we can use a type $V$ satisfying the following equality $\Id~{\UU_{\beta}}~V~(\Sigma_{X:\UU_{\alpha}}X\rightarrow V)$.
This equation is only well-typed modulo the constraint $\alpha<\beta$.

We can define in our system a constant
$$
c~=~\lam{\alpha~\beta}\lam{\alpha<\beta}\lambda_{Y:\UU_{\beta}}\Id~{\UU_{\beta}}~Y~ (\Sigma_{X:\UU_{\alpha}}X\rightarrow Y)~~:~~
   [\alpha~\beta][\alpha<\beta](\UU_{\beta} \rightarrow \UU_{\beta^+})
$$

   This is because $\Sigma_{X:\UU_{\alpha}}X\rightarrow Y$ has type $\UU_{\beta}$ in the context

   $$\alpha:\Level,~\beta:\Level,~\alpha<\beta,~Y:\UU_{\beta}$$

   We can further instantiate this constant $c$ on two levels $l$ and $m$, and this will be of type
   $$[l<m](\UU_{m} \rightarrow \UU_{m^+})$$
   and this can only be used further if $l<m$ holds in the current
   context\footnote{It is interesting to replace $\Id~\UU_\beta$ in the
   definition of
$c$ above by $\Eq$. We leave it to the reader to verify the
following typing, for which no constraint is needed:
$$
c'~=~\lam{\alpha~\beta}\lambda_{Y:\UU_{\beta}}~\Eq~Y~ (\Sigma_{X:\UU_{\alpha}}X\rightarrow Y)~~:~~
   [\alpha~\beta](\UU_{\beta} \rightarrow \UU_{\beta\vee\alpha^+})
$$}.

\medskip

In the current system of Agda, the constraint $\alpha<\beta$ is represented indirectly by
writing $\beta$ on the form $\gamma\vee \alpha^+$ and $c$ is defined as
$$
c~=~\lam{\alpha~\gamma}\lambda_{Y:\UU_{\alpha^+\vee\gamma}}\Id~{\UU_{\alpha^+\vee\gamma}}~Y~ (\Sigma_{X:\UU_{\alpha}}X\rightarrow Y)~~:~~[\alpha~\gamma]  (\UU_{\alpha^+\vee\gamma} \rightarrow \UU_{\alpha^{++}\vee\gamma^+})
$$
   which arguably is less readable.

\medskip

In general, if we  build a term $t$ of type $A$ in a context using labels $\alpha_1,\dots,\alpha_m$
and constraint $\psi$ and variables $x_1:A_1,\dots,x_n:A_n$ we can introduce a constant
$$
c~=~ \lam{\alpha_1~\dots~\alpha_m}\lam{\psi}\lambda_{x_1~\dots~x_n}t ~:~
[\alpha_1~\dots~\alpha_m][\psi]\Pi_{x_1:A_1~\dots~x_n:A_n}A
$$
We can then instantiate this constant $c~l_1~\dots~l_m~u_1~\dots~u_n$, but only if the levels
$l_1~\dots~l_m$ satisfy the constraint $\psi$.

We remark that Voevodsky's system \cite{VV} has no constraint-indexed products and no associated application operation, and instantiation of levels is only a meta-level operation. Sozeau and Tabareau \cite{SozeauTabareau:coq} do not have constraint-index products either. However, they do have a special operation
 for instantiating universe-polymorphic constants defined in the global environment.

\begin{remark}
Let's discuss some special cases and variations.

First, it is possible not to use
level variables at all, making the semilattice empty,
in which case the type theory defaults to one without universes
as presented in Section~\ref{sec:basic}.

Second, one could have exactly one level variable in the context.
Then any constraint would either be a loop or trivial.
In the latter case, the finitely presented semilattice
is isomorphic to the natural numbers with successor and $\max$.
Still, we get some more
expressivity than the type theory in Section~\ref{sec:external} since
we can express universe polymorphism in one variable.

Third, with arbitrarily many level variables but not using constraints
we get the type theory in Section~\ref{sec:internal}.

Fourth, we could add a bottom element, or empty supremum, to the semilattice.
Without level variables and constraints, the finitely presented semilattice
is isomorphic to the natural numbers with successor and $\max$
and we would get the type theory in Section~\ref{sec:external}.
We would also get a first universe.
(Alternatively, one could have one designated level variable
$0$ and constraints $0\leqslant\alpha$
for all level variables $\alpha$.)

Fifth, we note in passing that the one-point semilattice
with $\_^+$ has a loop.
\end{remark}

\section{Related work}\label{sec:related}

We have already discussed both Coq's and Agda's treatment of universe polymorphism in the introduction, including the work of Huet, Harper and Pollack, Courant, Herbelin, and Sozeau and Tabareau, as well as of Voevodsky. In this section we further
discuss the latter two, as well as some recent related work.

\paragraph*{Lean}

One can roughly describe the type system of Lean \cite{moura:lean,Carneiro19} as our current type system
where we only can declare constants of the form
$c~=~\lam{\alpha_1~\dots~\alpha_n}M~~:~~[\alpha_1~\dots~\alpha_n]A$
where there are no new level variables introduced in $M$ and
$A$.

\paragraph*{Voevodsky} One of our starting points was the 79 pp.\ draft \cite{VV} by Voevodsky, where type theories are parametrized by a fixed but arbitrary finite set of constraints over a given finite
set $\Fu$ of \emph{u-level variables}. A \emph{u-level expression} \cite[Def. 2.0.2]{VV} is either a numeral,
or a variable in $\Fu$, or an expression of the form $M+n$ with $n$
a numeral and $M$ a u-level expression, or of the form $\max(M_1,M_2)$
with $M_1,M_2$ u-level expression. A \emph{constraint} is an equation
between two u-level expressions. Given the finite set of constraints,
$\AFu$ is the set of assigments of natural numbers to variables
in $\Fu$ that satisfy all constraints.

The rules 7 and 10 in \cite[Section 3.4]{VV} define how to use constraints:
two types (and, similarly, two terms) become definitionally equal
if, for all assignments in $\AFu$, the two types become \emph{essentially}
syntactically equal after substitution of all variables in $\Fu$ by
their assigned natural number. For example, the constraint
$\alpha < \beta$ makes $\UU_\beta$ and $\UU_{\max(1,\beta)}$
definitionally equal.

For decidability, Voevodsky refers in the proof of
\cite[Lemma 2.0.4, proof]{VV} to Presburger Arithmetic,
in which his constraints can easily be expressed.\footnote{%
For this it seems necessary to also require that $\AFu$
is defined by a \emph{finite} set of constraints.}
This indeed implies that definitional equality is decidable, even
``in practice [...] expected to be very easily decidable i.e.\
to have low complexity of the decision procedure''
\cite[p.\ 5, l.\ -13]{VV}.
The latter is confirmed by \cite{bezem-coquand:lattices}.

The remaining sections of \cite{VV} are devoted to extending the
type theory with data types, $W$-types and identity types,
and to its metatheory.

We summarize the main differences between our type theories
and Voevodsky's as follows.
In \cite{VV}, u-levels are natural numbers, even though u-level
expressions can also contain u-level variables, successor and maximum.
Our levels are elements of an abstract sup-semilattice with a successor
operation. In the abstract setting, for example,
$\alpha\vee\beta=\alpha^+$ does not imply $\beta=\alpha^+$,
whereas in \cite{VV} it does.
In \cite{VV}, constraints are introduced, once and for all,
at the level of the theory. In our proposal they are introduced
at the level of contexts.
There are no level-indexed products and no constraint-indexed products in \cite{VV}.
We also remark that Voevodsky's system is Tarski-style and has cumulativity (rules 29 and 30 in \cite[Section 3.4]{VV}). Our system is also Tarski-style, but we present a Russell-style version in Appendix~\ref{appendix:2}. We present rules for cumulativity in Appendix~\ref{appendix:1}.

\paragraph*{Sozeau and Tabareau}

In Sozeau and Tabareau's \cite{SozeauTabareau:coq} work on universe polymorphism in the Coq tradition, there are special rules for introducing universe-monomorphic and universe-polymorphic constants, as well as a rule for instantiating the latter.
However, their system does not include the full explicit universe polymorphism provided by level- and constraint-indexed products.
In our system, with explicit universe polymorphism, we can have a uniform treatment of definitions, all of the form
$$ c : A = t$$
where $A$ is a type and $t$ a term of type $A$, and these definitions can be local as well.

The constraint languages differ: their constraints are equalities or (strict) inequalities between level variables, while ours are equalities between level expressions generated by the supremum and successor operations.

Furthermore, they consider cumulative universe hierarchies  \`a la Russell,
while our universes are \`a la Tarski and we consider both non-cumulative
(like Agda) and cumulative versions.

One further important difference is that their system has been completely implemented and tested on
significant examples, while our system is at this stage only a proposal. The idea would be that the
users have to declare explicitly both universe levels {\em and} constraints. The Agda implementation
shows that it works in practice to be explicit about universe levels, and we expect that to be
explicit about constraints will actually simplify the use of the system,
but this has yet to be tested in practice.
Recently, Coq has been extended to support universes and constraint
annotations from entirely implicit to explicit. Moreover, our level- and
constraint-indexed products can to some extent be simulated by using
Coq's module system \cite{coq:univpoly}.

\paragraph*{Assaf and Thir\'e}

Assaf \cite{Assaf14} considers an alternative version of the calculus of
constructions where subtyping is explicit. This new system avoids problems related to coercions and dependent types by using the Tarski style
of universes and by introducing additional equations to reflect equality. In particular he adds an explicit cumulativity map $\T^0_1 : \UU_0 \to \UU_1$. He argues that ``full reflection'' is necessary to achieve the expressivity of Russell style. He introduces the explicit cumulative calculus of constructions (CC$\uparrow$) which is closely related to our system of externally indexed Tarski style universes.
This is analysed further in the PhD thesis of F. Thir\'e \cite{Thire20}.

\section{Conjectures and future work}\label{sec:future}

Canonicity and normalization have been proved for a type theory with an external tower of universes \cite{coquand:tcs2019}. We conjecture that these proofs can be modified to yield proofs of analogous properties (and their corollaries) for our type theories in Section~\ref{sec:internal} and \ref{sec:constraints}.
In particular, decidability of type checking should follow using
\cite{bezem-coquand:lattices}.

\paragraph*{Acknowledgement}
The authors are grateful to the anonymous referees for useful feedback,
and to Matthieu Sozeau for an update on the current state
of universe polymorphism in Coq.
We acknowledge the support of the Centre for Advanced Study (CAS)
at the Norwegian Academy of Science and Letters
in Oslo, Norway, which funded and hosted the research project Homotopy
Type Theory and Univalent Foundations during the academic year 2018/19.

\bibliography{refs-lipics}

\newpage

\appendix

\section{Formulation with cumulativity} \label{appendix:1}

We introduce an operation $\T_{l}^{m}(A):\UU_{m}$ if $A:\UU_{l}$
and $l\leqslant m$ (i.e., $m = l\vee m$).\footnote{%
Recall that the equality of universe levels is the one of
sup-semilattice with the $\_^+$ operation.}

We require $\T_{m}(\T_{l}^{m}(A)) = \T_{l}(A)$. Note that this
yields, e.g., $a:\T_{m}(\T_{l}^{m}(A))$ if $a:\T_{l}(A)$.
We also require $\T_{l}^{m}(\NN^{l}) = \NN^{m}$ ($l\leqslant m$),
and $\T_{l}^{m}(\UU_{k}^l) = \UU_{k}^{m}$ ($k<l\leqslant m$),
as well as $\T_{l}^m(A) = A$ ($l=m$)
and $\T_{m}^n(\T_{l}^m(A)) = \T_l^n(A)$ ($l\leqslant m\leqslant n$),
for all $A:\UU_l$.

We can then simplify the product and sum rules to
$$
\frac{A:\UU_{l}~~~~~~B:\T_{l}(A)\rightarrow \UU_{l}}
     {\Pi^{l} A B:\UU_{l}}~~~~~~~~~
\frac{A:\UU_{l}~~~~~~B:\T_{l}(A)\rightarrow \UU_{l}}
     {\Sigma^{l} A B:\UU_{l}}~~~~~~~~~
$$
with conversion rules
$$
\T_{l}~(\Pi^{l} A B) = \mypi{x}{\T_{l}(A)}{ \T_{l}(B~x)}~~~~~~~
\T_{l}~(\Sigma^{l} A B) =  \mysig{x}{\T_{l}(A)}{ \T_{l}(B~x)}~~~~~~~
$$
and
$$
\T_{l}^{m}~(\Pi^{l} A B) = \Pi^{m} \T_{l}^{m}(A) (\mylam {x}{\T_{l}(A)}{\T_{l}^{m}(B~x)})~~~~~~
\T_{l}^{m}~(\Sigma^{l} A B) = \Sigma^{m} \T_{l}^{m}(A) (\mylam {x}{\T_{l}(A)}{\T_{l}^{m}(B~x)})~~~~~~
$$

Recall the family $\Id^l(A,a,b):\UU_l$ for $A:\UU_l$
and $a:\T_l(A)$ and $b:\T_l(B)$, with
judgemental equality $\T_l(\Id^l(A,a,b)) = \Id(\T_l(A),a,b)$.
We add the judgmental equalities $\T_l^m(\Id^l(A,a,b)) =
\Id^m(\T_l^m(A),a,b)$; note that $a$ and $b$ are well-typed since
$\T_{m}(\T_{l}^{m}(A)) = \T_{l}(A)$.

\medskip

Example. Recall the type $\Eq^{l,l}(A,B):\UU_l$ for $A$ and $B$ in $\UU_l$,
with judgmental equality
$\T_l(\Eq^{l,l}(A,B)) = \Equiv(\T_l(A),\T_l(B))$.
For $m>l$, a consequence of univalence
for $\UU_m$ and $\UU_l$ is that we can build an element of the type
$$
\Id(\UU_m,\Eq^{m,m}(\T^m_l(A),\T^m_l(B)),\Id^m(\UU^m_l,A,B)).
$$

\section{Notions of model and formulation \`a la Russell} \label{appendix:2}

\paragraph*{Generalised algebraic presentation}

In a forthcoming paper, we plan to present some generalised algebraic theories of level-indexed categories with families with extra structure.  The models of these theories provide suitable notions of model of our type theories with level judgments. Moreover, the theories presented in this paper are initial objects in categories of such models.
\begin{remark} \label{app:annotation}
As explained
in \cite{streicher:semtt}, in order to see the theories in this paper as presenting {\em initial} models,
it is enough to use a variation where application $c~a:B(a/x)$ for $c:\mypi{x}{A}B$ and $a:A$
is annoted by the type family $A,B$ (and similarly for the pairing operation). If the theories satisy the normal form property, it can then be shown that also the theories without annotated application are initial.
\end{remark}

\paragraph*{Russell formulation}
Above, we presented type theories with universe level judgments {\em \`a la Tarski}. There are alternative formulations
{\em \`a la Russell} (using the terminology introduced in
\cite{martinlof:padova} of
universes). One expects these formulations to be equivalent to the Tarski-versions, and thus also initial models. For preliminary results in this
direction see \cite{Assaf14,Thire20}.

With this formulation, the version without cumulativity becomes

$$
\frac{A:\UU_{n}}{A~\type}
$$
$$
\frac{A:\UU_{n}~~~~~~B:\UU_m(x:A)}
     {\mypi{x}{A}{B}:\UU_{n\vee m}}~~~~~~~~~
\frac{A:\UU_{n}~~~~~~B:\UU_m(x:A)}
     {\mysig{x}{A}{B}:\UU_{n\vee m}}~~~~~~~~~
$$
$$\frac{l~\Level}{\NN:\UU_{l}}$$
$$
\frac{A:\UU_n~~~~~~a_0:A~~~~~~a_1:A}{\Id(A,a_0,a_1):\UU_n}
$$
$$
\frac{l<n}{{\UU_l}:\UU_{n}}
$$

\medskip

For the version with cumulativity, we add the rules
$$
\frac{A:\UU_{l}~~~~~~l\leqslant n}{A:\UU_{n}}
$$
and the rules for products and sums can be simplified to
$$
\frac{A:\UU_{n}~~~~~~B:\UU_n~(x:A)}
     {\mypi{x}{A}{B}:\UU_{n}}~~~~~~~~~
\frac{A:\UU_{n}~~~~~~B:\UU_n~(x:A)}
     {\mysig{x}{A}{B}:\UU_{n}}~~~~~~~~~
$$

     For $m>l$ the consequence of univalence for $\UU_m$ and $\UU_l$
     mentioned in Appendix~\ref{appendix:1} can now be written
     simply as
     $$\Id(\UU_m,\Equiv(A,B),\Id(\UU_l,A,B)).$$

\begin{remark}\label{notelementary}
  In the version \`a la Tarski, with or without cumulativity, terms have unique types, in the sense that if $t : A$ and $t : B$ then $A = B$, by induction on $t$. But for this to be valid, we need to annotate application as discussed in Remark~\ref{app:annotation}.
  Even with annotated application, the following property is not elementary: if $\UU_n$ and $\UU_m$ are convertible then $n$ is equal to $m$. This kind of property is needed for showing the equivalence between the Tarski and the Russell formulation.
\end{remark}

\begin{remark} \label{uniqueness:without:cumulativity}
If, in a system without cumulativity, we extend our system of levels with a least level 0, then if we restrict $\NN$ to be of type $\UU_0$, and $\UU_n$ to be of type $\UU_{n+1}$ then well formed terms have unique types.
\end{remark}

\begin{remark}
  It should be the case that the above formulation \`a la Russell presents the initial CwF with extra extructure for the standard type formers and a hierarchy of universes, but the proof doesn't seem to be trivial, due to Remark~\ref{notelementary}.
\end{remark}

\section{Revision and extension} \label{appendix:3}

The type theory in \cref{sec:constraints} has received some justified,
and very constructive, criticism. Let's explain the problem first,
and then discuss some solutions.

As an example, consider the inferrable typing
$\alpha,\beta \vdash [\alpha<\beta] \NN~\type$.
Substituting $\alpha$ for $\beta$ gives the typing
$\alpha \vdash [\alpha<\alpha] \NN~\type$.
However, the latter typing cannot be inferred.
The reason is that $\alpha<\alpha$ is a loop
and cannot occur in any context, as context extension
is only allowed if the resulting context is loop-free.
Hence the formation rule for constraint-indexed products
(the first rule in \cref{fig:restriction}) cannot be applied.

As a consequence of the above example,
the type theory in \cref{sec:constraints}
fails to have some usual, and very desirable, metatheoretic properties,
such as substitution invariance and subject reduction.
(Still, we believe our type theory to be consistent.)
There are several ways to solve this problem.
Let's first take a closer look at substitutions.

Recall that contexts in \cref{sec:constraints}, as far as the levels
are concerned, are finitely presented \emph{loop-free} lattices in the
sense of \cite{bezem-coquand:lattices}. Level substitutions
are lattice homomorphisms between these contexts.
The substitution $\beta/\alpha : (\alpha) \to (\alpha,\beta)$ is allowed,
leading to the example above. The resulting type $[\alpha<\alpha] \NN$
can be considered to be an `empty' product. The substitution
$\beta/\alpha : (\alpha,\alpha<\alpha) \to (\alpha,\beta,\alpha<\beta)$
is not allowed, since $\alpha<\alpha$ is a loop.

One possible solution is to present the system with explicit substitutions,
with suitable judgemental equality rules. We can then be careful when such a substitution makes a constraint product empty,
replacing this empty product by a unit type
(and replacing the corresponding abstraction by the unique element
of this unit type). More precisely,
whenever $\sigma:\Delta\rightarrow\Gamma$ is a
substitution and $\psi\sigma$ introduces a loop
for $\Delta$, which is decidable, we add the
conversion that $([\psi]A)\sigma$ is equal to the unit type.

As an example, if $\sigma$ is the weakening
$(\alpha,\beta,\alpha = \beta)\to (\alpha,\beta)$ and $A$ is a
type in the context $\alpha,\beta,\alpha<\beta$, then $[\alpha < \beta]A$
is a type in the context $\alpha,\beta$
and we add the conversion that $([\alpha<\beta]A)\sigma$ is equal to the
unit type, since $(\alpha<\beta)\sigma = \alpha<\beta$
introduces a loop for the context $\alpha,\beta,\alpha = \beta$.

We refrain from elaborating the above solution in detail here, as it will complicate the syntax significantly. Instead we will present
in the next section a short and elegant solution suggested
by Georges Gonthier, for which we are very grateful.

\paragraph*{Allowing loops in contexts}
\def\gg{{\emptyset}}

First we liberalize the context extension operation $\Gamma,\psi$ in
\cref{sec:constraints} by dropping the second condition,
that the extended context $\Gamma,\psi$ is loop-free.

Recall that a level expression $l$ is a {loop}
if $l<l$ is implied by the constraints in the context $\Gamma$,
and that the existence of such a loop is decidable. Thus we can use
the judgement $\Gamma\vdash l<l~\valid$ for expressing that $\Gamma$
yields a loop.

We assume that our type theory has an empty type $\gg$.
The idea of adding the following rules is to collapse the type
theory in contexts that yield a loop.

\begin{figure}[h!]
$$
\frac{A~\type~~~l<l~\valid}{\gg = A}~~~~~~~~~~~~~~~~~~~~
\frac{t:A~~~l<l~\valid}{\gg = t : A}
\belowdisplayshortskip 1pt
$$
 \caption{Rules for collapsing}%
  \label{fig:collapsing}
\end{figure}
Note that we can infer $\Gamma \vdash \gg : \gg$ if $\Gamma$ yields a loop.
(The lattice presented by the level
variables and constraints in $\Gamma$ can still be non-trivial even
if $\Gamma$ yields a loop.)
In this solution no explicit substitutions are needed.

\end{document}